\documentclass[aps,prd,reprint,superscriptaddress]{revtex4-1}

\usepackage{bm,graphics,graphicx}
\usepackage{color}
\newcommand{\ds}{\displaystyle}

\begin{document}


\title{Spacetime rotation-induced Landau quantization}


\author{Kohkichi Konno}
\email[]{kohkichi@gt.tomakomai-ct.ac.jp}
\affiliation{Department of Natural and Physical Sciences,
  Tomakomai National College of Technology, 
  Tomakomai 059-1275, Japan}
\affiliation{Center of Education and Research for
 Topological Science and Technology, 
 Hokkaido University, Sapporo 060-8628, Japan}

\author{Rohta Takahashi}
\email[]{rohta@riken.jp}
\affiliation{Department of Natural and Physical Sciences,
  Tomakomai National College of Technology, 
  Tomakomai 059-1275, Japan}
\affiliation{High Energy Astrophysics Laboratory, 
 The Institute of Physical and Chemical Research (RIKEN), 
 Saitama 351-0198, Japan}


\date{\today}

\begin{abstract}
We investigate non-inertial and gravitational effects on quantum 
states in electromagnetic fields and 
present the analytic solution for energy eigenstates for
the Schr\"odinger equation including non-inertial, gravitational 
and electromagnetic effects. We find that in addition to 
the Landau quantization the rotation of spacetime itself 
leads to the additional quantization, 
and that the energy levels for an electron are 
different from those for a proton 
at the level of gravitational corrections. 
\end{abstract}

\pacs{04.20.Cv, 03.65.Ge, 04.62.+v}

\maketitle

%
{\it Introduction}.---
In many quantum systems, gravitational interaction
is usually neglected because of the weakness of the interaction. 
Hence, gravitational effects on quantum systems
remain to be fully elucidated. At present, 
such circumstance may be an obstacle to understand 
the interplay between the quantum theory 
and the gravitational theory.
As for verifications of gravitational effects on quantum systems, 
several experiments have been conducted.
Colella, Overhauser and Werner \cite{cow} for the first time 
experimentally showed an physical phenomenon involving 
both the Plank constant $\hbar$ and the gravitational
constant $G$ 
by using a neutron interferometer elegantly. Since then, 
ingenious experiments using neutrons \cite{wsc,bw,vvn} 
or atoms \cite{atoms}
have been conducted to reveal gravitational or
non-inertial effects on quantum systems. 

On the other hand, electromagnetic fields are ubiquitous 
in the universe. Around magnetized compact objects such 
as magnetized neutron stars and magnetors, the couplings 
between gravitational effects, quantum effects and 
electromagnetic effects will come into play.  
Actually, signatures of Landau quantization 
in X-ray cyclotron absorption lines were observed 
on a neutron star surface \cite{bcdm} where the gravitational effect 
are much stronger than that on the Earth. 
While non-inertial and gravitational effects 
on quantum systems in unmagnetized circumstances have 
been well studied theoretically so far
\cite{spin0,spin,lf,kkf,ab,kk},
there are only a few reports 
\cite{cvz,kw,bakke,ac,mism} about those effects 
in magnetized circumstances in the literature.

In this paper, we investigate non-inertial and gravitational effects 
on quantum systems 
in electromagnetic fields by solving 
the Schr\"{o}dinger equation seriously 
for non-relativistic magnetized matter
in slowly rotating Kerr spacetime, 
and find the analytic solution for the quantum states 
of a charged particle including 
non-inertial, gravitational, and electromagnetic effects
for the first time, in which
we neglect the effect of the intrinsic spin 
of a particle \cite{spin,kk,mism}.

{\it Spacetime metric}.---
First of all, we discuss the metric around a rotating star.
We assume that the rotational axis is aligned with the $z$-axis.
In this paper, we explicitly use the gravitational constant $G$ and 
the speed of light $c$ for later conveniences.
The spacetime metric is approximated by the 
slow rotation limit of the Kerr metric
\begin{eqnarray}
 ds^2 & = &\left(1-\frac{2M_*}{r}\right) c^2 dt^2 + \frac{4M_*a}{r} \sin^2 \theta \; c dt d\phi
\nonumber\\&&
- \frac{dr^2}{1-2M_*/r}
-r^2\left( d\theta^2 + \sin^2 \theta d\phi^2 \right) ,~~~
\end{eqnarray}
where $M_*=GM/c^2$,  
$M$ is the mass of the star and 
$a$ is the Kerr parameter and is considered to be
small throughout this paper.
After the coordinate transformation
$(x,y,z) = (\rho \sin\theta \cos\phi , \
\rho \sin\theta \sin\phi , \ \rho \cos\theta )$,
where
$\rho= \left( r-M_*+\sqrt{r^2- 2M_* r} \right)/2$,
we derive 
\begin{eqnarray}
 ds^2 & = & {\cal F}^2 {\cal G}^{-2} c^2 dt^2 
  + \frac{4M_*a}{\rho^3} {\cal G}^{-2} 
  \left( xdy-ydx \right) c~dt \nonumber \\
 &&
  - {\cal G}^{4} 
  \left( dx^2 + dy^2 + dz^2 \right) ,
\end{eqnarray}
where ${\cal F} \equiv  1 - M_*/(2\rho)$ and
${\cal G} \equiv  1 + M_*/(2\rho)$.
Furthermore we consider the coordinate transformation to 
the rotating frame on the stellar surface, i.e.,
$(x,y,z) = ( x' \cos \Omega t - y' \sin \Omega t ,
x' \sin \Omega t + y' \cos \Omega t , z' )$,
where $\Omega$ is the angular velocity of the rotating star.
Dropping the prime after the transformation,
we obtain
\begin{eqnarray}
\label{eq:metric}
 ds^2 & = & \left[ {\cal F}^2 {\cal G}^{-2} c^2 
 + \frac{4M_* c a}{\rho^3} {\cal G}^{-2}  \Omega \left( x^2 + y^2 \right) 
 - {\cal G}^{4} \Omega^2 \right. \nonumber \\
&& 
  \times \left( x^2 + y^2 \right)  \bigg] dt^2 
  + 2 \left( 
  \frac{2M_* c a}{\rho^3} {\cal G}^{-2}-{\cal G}^{4}\Omega 
  \right) \nonumber \\
&& \times
 \left( xdy-ydx \right) dt
  - {\cal G}^{4} 
  \left( dx^2 + dy^2 + dz^2 \right) .
\end{eqnarray}
Equation (\ref{eq:metric}) provides the spacetime metric which 
is described by an observer on the surface of a rotating star. 

\begin{figure}
 \includegraphics[width=0.45 \textwidth]{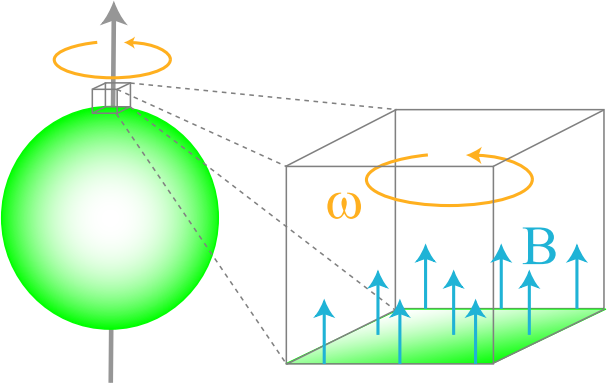}
 \caption{\label{fig1}Illustration of the polar region 
          of a rotating star. We assume that the magnetic axis 
          is aligned with the rotational axis.}
\end{figure}

{\it The Schr\"odinger equation with general relativistic 
corrections}.---Let us obtain the Schr\"odinger equation 
with general relativistic corrections 
from the Klein-Gordon equation \cite{kkf}.
Our approach is validated only when we neglect the intrinsic spin.
The Klein-Gordon equation for a massive scalar field $\phi$
in the presence of an electromagnetic field is given by 
\cite{cvz,kw,ac}
\begin{equation}
\label{eq:kg}
 \left[ g^{\mu\nu} \left( \nabla_{\mu} 
 - \frac{iq}{\hbar} A_{\mu} \right)
  \left( \nabla_{\nu} 
 - \frac{iq}{\hbar} A_{\nu} \right)
 + \frac{m^2c^2}{\hbar^2} \right] \phi = 0 ,
\end{equation}
where $m$ is the mass of the field $\phi$, 
$q$ is the electric charge, $g^{\mu\nu}$ is the metric, 
and $\nabla_{\mu}$ denotes the covariant derivative. 
Now we focus on the polar region of the rotating star
(see FIG.~\ref{fig1}).
The magnetic field is approximated by a uniform magnetic field
in the polar region. Here we assume that the magnetic field
lines are aligned with the rotational axis. 
This special configuration enables us to obtain 
the analytic solution for the field.
Thus we take 
the electromagnetic 4-potential as
$A_{\mu} = \left( 0, - By/2 , Bx/2 , 0 \right)$.
To derive the Schr\"odinger equation, 
we expand the field $\phi$ as
$\phi \left( t, x,y,z \right)
 = \psi \left( t, x,y,z \right){\rm exp}[- i (mc^2/\hbar) t] $.
From Eq.~(\ref{eq:kg}), up to $O\left( c^{-1} \right)$,
we obtain 
\begin{eqnarray}
 i\hbar \frac{\partial}{\partial t} \psi 
 & = & \left[ - \frac{\hbar^2}{2m} \nabla^2 - \frac{GMm}{\rho} 
  - \left( \frac{qB}{2m} 
  + \varpi (\rho ) \right) L_{z}
 \right. \nonumber \\
 && \left. 
  +  \left( \frac{q^2 B^2}{8m} 
  + \frac{qB}{2} \varpi (\rho )\right) 
  \left( x^2+y^2 \right)
  \right] \psi , 
\end{eqnarray}
where $\nabla^2 \equiv \partial_{x}^2 + \partial_{y}^2 + \partial_{z}^2$, 
$L_{z} \equiv -i\hbar \left( x \partial_{y} - y \partial_{x} \right)$, 
and $\varpi (\rho ) \equiv \Omega - 2GMa/(c\rho^3)$.
To take the origin of the $z$ axis at the surface, 
we transform $z$ as $z \rightarrow R+z$, where 
$R$ is the radius of the star. 
Here we have $x, y, z \ll R$ for the polar region and 
derive
$\rho \simeq R \left( 1+ zR^{-1} \right)$. 
For energy eigenstates, we obtain
\begin{eqnarray}
\label{eq:master-eq0}
 E \psi 
 & = & \left[ - \frac{\hbar^2}{2m} \nabla^2 +mU + mgz 
  - \left( \frac{qB}{2m} 
  + \varpi (R) \right) L_{z} \right.
  \nonumber \\
 && \left. 
  +  \left( \frac{q^2 B^2}{8m} 
  + \frac{qB}{2} \varpi (R) \right) 
  \left( x^2+y^2 \right)
  \right] \psi .
\end{eqnarray}
where $E$ is the energy, 
$U\equiv -GM/R$ is the gravitational
potential,
and $g\equiv GM/R^2$ is the gravitational acceleration.
Here we have neglected the term 
proportional to $az/R^4$ in $\varpi$.
Equation (\ref{eq:master-eq0}) governs quantum states 
on the surface of a rotating star.

{\it Quantum states on the surface of a rotating star}.---
We discuss quantum states on the stellar surface.
To solve Eq.~(\ref{eq:master-eq0}) for $\psi$, 
we assume the separation of variables as
$\psi \left( x,y,z \right) = F\left( x,y \right) G(z)$,
where functions $F$ and $G$ are introduced.
From Eq.~(\ref{eq:master-eq0}), we can 
derive the differential equation for $F$ and 
that for $G$ in the cylindrical coordinates $( r,\theta , z)$ 
\begin{eqnarray}
\label{eq:xy-direction}
 \lefteqn{
 \left( E -mU - K \right) F( r ,\theta )
 } \nonumber \\
 & = &   \left[ 
  - \frac{\hbar^2}{2m} \left( \partial_{r}^2
  + \frac{1}{r} \partial_{r} \right) 
  + \frac{\hat{p}_{\theta}^2}{2mr^2} 
  - \left( \frac{qB}{2m} 
  + \varpi (R) \right)
  \hat{p}_{\theta}
  \right.\nonumber \\
 && \left.
  +  \left( \frac{q^2 B^2}{8m} 
  + \frac{qB}{2} \varpi (R) \right) r^2
  \right] F( r , \theta ) , 
\end{eqnarray}
\begin{equation}
\label{eq:z-direction}
 K G(z) = \left[ - \frac{\hbar^2}{2m} \frac{d^2}{dz^2}
 +mgz \right] G(z) ,
\end{equation}
where $\hat{p}_{\theta} \equiv - i\hbar \partial_{\theta}$, 
and $K$ is a constant introduced by the method of separation 
of variables. 
When we consider a neutron star, we should recall that 
the electron capture process is dominant inside a neutron star.
When we take account of the potential inside a
neutron star, we should replace the latter equation 
(\ref{eq:z-direction}) with the equation 
\begin{equation}
\label{eq:z-direction2}
 K G(z) = \left[ - \frac{\hbar^2}{2m} \frac{\partial^2}{\partial z^2}
 + V_{\rm eff}(z) \right] G(z) ,
\end{equation}
where the effective potential 
$V_{\rm eff} (z)$ is assumed to be 
\begin{equation}
 V_{\rm eff} (z) = \left\{ 
 \begin{array}{cc}
  mgz & (z>0) \\ \infty & (z\le 0) 
 \end{array}
 \right. .
\end{equation}
Here the form of $V_{\rm eff}$ for $z\le 0$ may be somewhat ideal. 
(See {\it Discussion \& Conclusion} for anticipation 
of more realistic cases.)
Thus, we can find quantum states on 
the neutron star by solving Eqs.~(\ref{eq:xy-direction})
and (\ref{eq:z-direction2}).

First we discuss Eq.~(\ref{eq:z-direction2}) for the wave function
in the $z$-direction. This equation was already 
investigated in \cite{lf,sakurai}. 
The solution of Eq.~(\ref{eq:z-direction2})
is given by the Airy function 
\begin{equation}
\label{eq:Gz}
 G(z) = {\rm Ai} \left( \left( \frac{2m^2 g}{\hbar^2} \right)^{\frac{1}{3}}
  \left( z-\frac{K}{mg} \right) \right) .
\end{equation}
Here $K$ is quantized due to the boundary condition 
at $z=0$, i.e., $G(0)=0$, as
$
 K_{n} = \hbar \omega_{\perp} (m) \lambda_{n},
$
where $n=0,1,2,\cdots$, 
$\omega_{\perp} (m) \equiv 
\left( m g^2 / (2\hbar) \right)^{1/3}$
and
$\lambda_{n}$ denotes the zero points of the Airy function, 
i.e., ${\rm Ai}(-\lambda_{n}) =0$.
Therefore, the wave function in the $z$-direction
is given by Eq.~(\ref{eq:Gz}) with quantized energy 
$K=K_{n}$.

Next, we discuss Eq.~(\ref{eq:xy-direction}) in the $xy$-plane.
Let us take eigenstates for $\hat{p}_{\theta}$, i.e., 
$F( r, \theta ) 
  = {\rm exp}[i (p_{\theta}/\hbar) \theta] f(r) $,
where $f$ is a function of $r$ only. From 
Eq.~(\ref{eq:xy-direction}), we derive 
\begin{equation}
 \left[ \frac{d^2}{dr^2} + \frac{1}{r} \frac{d}{dr} 
  - \beta^2 r^2 - \frac{p_{\theta}^2}{\hbar^2 r^2} + {\cal E} \right] f = 0 , 
\end{equation}
where 
\begin{eqnarray}
 \beta & = & \left( \frac{q^2 B^2}{4\hbar^2} 
  + \frac{mqB}{\hbar^2} \varpi (R) 
  \right)^{\frac{1}{2}} , \\
 {\cal E} & = & 
  \frac{2m}{\hbar^2}\left( E -mU - K_{n} \right) 
  +  \left( \frac{q B}{\hbar^2} 
  + \frac{2m}{\hbar^2} \varpi (R)
  \right) p_{\theta} .
\end{eqnarray}
Furthermore, we assume
$ f (r) =r^{\ell} e^{- \frac{\beta}{2} r^2} \tilde{f}(r) $,
where $\ell$ is defined as 
\begin{equation}
\label{eq:ell}
 \ell = \pm \frac{p_{\theta}}{\hbar}  \quad 
      \mbox{for} \quad  q = \pm e  ,
\end{equation}
where $e>0$ is the elementary charge.
When we adopt the variable $x=\beta r^2$, we derive
\begin{equation}
\label{eq:a-laguerre}
  \left[ x \frac{d^2}{dx^2} + \left\{ (\ell +1) - x \right\}
  \frac{d}{dx} 
  + \left( \frac{{\cal E}}{4\beta} -\frac{\ell +1}{2} 
  \right) \right] \tilde{f}(x) = 0 .
\end{equation}
This equation is equivalent to the 
confluent hypergeometric equation
$ x y^{\prime\prime}+ (\gamma -x)y^\prime-\alpha y = 0 $.
Hence the solutions of Eq.~(\ref{eq:a-laguerre})
are given by the confluent hypergeometric functions
in the form
\begin{eqnarray}
\label{eq:hypergeo}
 y \; = \; _{1}F_{1} \left( \alpha , \gamma ; x \right) 
 = 1 + \frac{\alpha}{\gamma} \frac{x}{1!}
  + \frac{\alpha \left( \alpha + 1 \right)}
  {\gamma \left( \gamma +1 \right)} \frac{x^2}{2!} + \cdots .
\end{eqnarray}
We now discuss the integrability condition of the wave function 
$F$. The integral of $F^{\ast} F$ is calculated as
\begin{eqnarray}
 \int_{0}^{2\pi} \!d\theta \int_{0}^{\infty} \!rdr
   F^{\ast} F
\label{eq:norm}
 & = & \frac{\pi}{\beta^{\ell +1}} 
   \int_{0}^{\infty} dx \: 
   x^{\ell} e^{-x} \left[ \tilde{f} (x) \right]^2 ,
\end{eqnarray}
where $F^{\ast}$ denotes the complex conjugate of $F$.
Thus, when the series in Eq.~(\ref{eq:hypergeo}) ends
at a finite order, the integral of Eq.~(\ref{eq:norm}) 
becomes finite. Therefore, to make the wave function 
integrable, the constant $\alpha$ in Eq.~(\ref{eq:hypergeo})  
must be zero or negative integers, 
i.e., $\alpha = -n'$, where $n'=0,1,2,\cdots $. 
In the same way, from Eq.~(\ref{eq:a-laguerre}), 
we obtain the condition
\begin{equation}
\label{eq:condition}
 \frac{{\cal E}}{4\beta} -\frac{\ell +1}{2} = n' .
\end{equation}
In this case, we can find integrable wave functions.
When the condition Eq.~(\ref{eq:condition}) is satisfied,
the solution of Eq.~(\ref{eq:a-laguerre}) is given by 
the associated Laguerre polynomials
$\tilde{f}(x)= L^{\ell}_{n'} (x)$.
Thus, we obtain 
\begin{equation}
\label{eq:xy-wavefunction}
 F(r,\theta) = e^{i \frac{p_{\theta} \theta}{\hbar}}
  r^{\ell} e^{- \frac{\beta}{2} r^2}
  L^{\ell}_{n'} \left( \beta r^2 \right) .
\end{equation}
For $q=\pm e$, Eq.~(\ref{eq:condition}) is approximately 
calculated as
\begin{equation}
\label{eq:xy-energy}
 E -mU -K_{n} \simeq
     \hbar \left( \frac{e B}{m} 
     \pm 2\varpi (R) \right) 
     \left( n'+\frac{1}{2}\right) .
\end{equation}
Equations (\ref{eq:xy-wavefunction}) and (\ref{eq:xy-energy})
describe the Landau quantization with general relativistic corrections
in the $xy$-plane.

Consequently, we obtain the wave function 
on the polar region 
\begin{eqnarray}
\label{eq:wave-func}
 \psi & = & {\cal A} r^{\ell} e^{- \frac{\beta}{2} r^2}
   L^{\ell}_{n'} \left( \beta r^2 \right) 
  {\rm Ai} \left( \left( \frac{2m^2 g}{\hbar^2} \right)^{\frac{1}{3}}
  z - \lambda_{n} \right) ,  
\end{eqnarray}
where ${\cal A}$ is a normalization factor, and  
$\ell$ is given by Eq.~(\ref{eq:ell}) and must be zero or positive.
The energy eigenvalues for $q=\pm e $ 
are given by 
\begin{eqnarray}
\label{eq:energy}
 E_{nn'} & \simeq &
  mU + \hbar \omega_{\perp} (m) \lambda_{n} 
+ \hbar \left( \frac{e B}{m} 
     \pm  2\varpi (R) \right) 
     \left( n'+\ds\frac{1}{2}\right) \!\!, \nonumber \\
\end{eqnarray}
where the positive sign corresponds to the case for a proton, and
the negative sign corresponds to the case for an electron.
Equations (\ref{eq:wave-func}) and (\ref{eq:energy})
provide the quantum states of the field $\psi$
on the surface of a rotating star.

\begin{table}
 \caption{\label{tab1} Vertical energy eigenstates for an electron 
  are compared with those for a proton. $m_{\rm e}$ and $m_{\rm p}$ 
  denote the mass for an electron and that for a proton, respectively.
  We adopt $M=M_{\odot}$ and $R=10$[km] for the estimates.} 
 \begin{ruledtabular}
 \begin{tabular}{cccc}
 $n$ & $\lambda_{n}$ & $K_{n}(m_{\rm e})/\hbar$[Hz] 
  & $K_{n}(m_{\rm p})/\hbar$[Hz] \\ \hline
 1 & 2.33811 & $4.61512 \times 10^9$ & $5.65135 \times 10^{10}$ \\
 2 & 4.08795 & $8.06908 \times 10^9$ & $9.88083 \times 10^{10}$ \\
 3 & 5.52056 & $1.08969 \times 10^{10}$ & $1.33435 \times 10^{11}$ \\
 4 & 6.78671 & $1.33961 \times 10^{10}$ & $1.64039 \times 10^{11}$ \\
 $\vdots$ & $\vdots$ & $\vdots$ & $\vdots$
 \end{tabular}
 \end{ruledtabular}
\end{table}

{\it Discussion \& Conclusion}.---We discuss physical consequences of 
the quantized states
with the general relativistic corrections. 
The energy states in Eq.~(\ref{eq:energy}) are characterized by 
two integers $n$ and $n'$. In Eq.~(\ref{eq:energy}), 
the first term is the gravitational potential and 
merely shifts the zero-point energy of the system. 
The second term denotes the energy levels in the vertical 
direction. The vertical energy levels depend on 
the mass $m$ only, not on the magnetic field $B$.
Thus the energy levels for a proton are different from 
those for an electron (see also TABLE \ref{fig1}).
The third term denotes the Landau quantization with 
the general relativistic correction. The sign in 
front of the correction $2\varpi$ depends on 
the charge of a particle. Thus the energy step
for a proton is different from that for an electron.  
Therefore, in principle, we could determine 
which particle comes into play on the surface
from the fine structure of the eigenstates due to 
the gravitational corrections.

It is worth noting that the Landau quantization 
caused by $\varpi$ survive in the limit of $B \rightarrow 0$. 
Therefore, the rotation of the spacetime cause 
the Landau quantization without magnetic fields.
This effect might be called {\em 
spacetime rotation-induced (or geometric) 
Landau quantization} 
(see also \cite{note1,note2}).

Next, 
we discuss observability of the quantum states 
discussed above for neutron stars.
Here it should be noted that we actually observe 
the energy that is subject to the gravitational redshift, i.e., 
$E_{\rm obs} = \gamma_{\rm red} E$, 
where $\gamma_{\rm red}$ is the factor for gravitational redshift.
For the vertical quantum levels, we can obtain the order estimates
$\gamma_{\rm red} \omega_{\perp} ( m_{\rm e} )
 \sim 10^{9} [\mbox{Hz}]$, 
$\gamma_{\rm red} \omega_{\perp} ( m_{\rm p} )
 \sim 10^{10} [\mbox{Hz}]$,
where $m_{\rm e}$ is the mass for an electron, and $m_{\rm p}$ 
is the mass for a proton. The first few eigen frequencies 
for the vertical energy levels are shown in TABLE \ref{tab1}.
In practice, the potential well $V_{\rm eff}$ would be broadened
out in the direction of $z<0$, and 
the intervals of the energy levels would become narrower.
When the magnetic field strength varies from 
$10^{8}[{\rm G}]$ to $10^{15}[{\rm G}]$, we derive 
the order estimates for cyclotron frequencies, 
$\gamma_{\rm red} e B / m_{\rm e} \sim 10^{15}\mbox{-}10^{22}  
 [\mbox{Hz}]$, 
$\label{eq:cycle-f2}
 \gamma_{\rm red} e B / m_{\rm p} \sim 10^{12}\mbox{-}10^{19}  
 [\mbox{Hz}]$.
For millisecond pulsars, we derive the order estimates 
of the rotational terms in $\varpi$ as
$2\gamma_{\rm red} \Omega \sim 10^{3} [\mbox{Hz}]$ and 
$\gamma_{\rm red} 4GMa/(cR^3) 
 = \gamma_{\rm red} 4GM / (c^2 R) \cdot J/(M R^2)
  \sim 10^2 [\mbox{Hz}]$.
Thus the cyclotron frequency is the most energetic for
neutron stars. The vertical energy step is the second
most energetic, and the general relativistic correction
to the Landau energy is the lowest.  
Hence, we can determine the quantity $B/m$ from 
the most energetic absorption lines that are almost given 
by the cyclotron frequency. 
While we can determine the mass $m$, in principle, 
from the gravitational corrections. 
Once we could detect the gravitational corrections 
from observations, 
we could determine the magnetic field strength itself. 
However, it would be difficult to detect the gravitational 
corrections at present. In general, the absorption lines 
are broaden 
by thermal, quantum and environmental effects \cite{rl}.
The probability of absorption in the vicinity of 
an absorption line would be proportional to the factor
$\exp (-|\Delta E| / (k_{\rm B} T))$, where $\Delta E$ is the 
energy difference from the absorption line, $k_{\rm B}$ 
is the Boltzmann constant, and $T$ is the temperature of the
environment. Since the surface of a neutron star typically 
has a temperature $T \sim 10^{6}[{\rm K}]$ \cite{meszaros}, 
the absorption line is broaden by the frequency 
width $\Delta \omega_{T} \sim k_{\rm B}T/\hbar \sim 10^{17}[{\rm Hz}]$. 
Thus the absorption lines below $\Delta \omega_{T}$ would be blurred. 
Although we could easily detect the cyclotron frequency above 
$\Delta \omega_{T}$, it would be difficult to detect 
the gravitational corrections owing to the thermal turbulence.
Nonetheless, the information of the gravitational corrections
is certainly hidden in the features of broaden absorption 
curves in spectra. 
This would be investigated in the future work. 

Although we have focused our attention on a neutron star,
the spacetime rotation-induced Landau quantization is 
universal phenomena. Thus the effect may be detectable 
with physical systems of ultra-low temperature such as 
superconductor in laboratories on the Earth \cite{note3}, 
rather than on a neutron star. Such an experimental verification 
would be awaited. 

\begin{acknowledgments}
We thank Profs.~S.~Tanda, S.~Mineshige, Y.~Eriguchi, 
T.~Tamagawa for continuous encouragements.
K.K.~thanks Prof. Y. Asano for useful conversation.
\end{acknowledgments}


\begin{thebibliography}{99}
\bibitem{note1}
  A similar effect of the rotational correction on 
  the quantization condition for a vortex 
  was discussed in \cite{kw}.
\bibitem{note2}
  An analog of Landau quantization for a neutral particle
  having magnetic moment in the presence of an electric field
  and a cosmic string was found in \cite{bakke}.
\bibitem{note3}
  When we consider the system located far from the pole, 
  we need to adopt the angular velocity component perpendicular 
  to the tangential plane of a stellar surface as $\varpi$.
  If the magnetic configuration is different from our assumption, 
  numerical investigation would be needed to derive accurate spectra.
\end{thebibliography}

\end{document}